\documentclass[%
floatfix,
reprint,%
onecolumn,
twocolumn,%
jcp,%
longbibliography,
]{revtex4-2}

\usepackage{graphicx}% Include figure files
\usepackage{dcolumn}% Align table columns on decimal point
\usepackage{bm} % bold math
\usepackage{color}

\usepackage[english]{babel}

\usepackage{amsmath,amsthm,latexsym,amssymb,amsfonts,epsfig}
\usepackage{mathtools}
\usepackage{mathrsfs} 

\usepackage[outer=1.6cm, inner=1.6cm, top=2.5cm, bottom=2.5cm]{geometry}  

\usepackage{multirow}

\usepackage{pifont}

\usepackage{ctable}

\usepackage{xcolor}
\definecolor{OliveGreen}{rgb}{0,0.6,0}
\definecolor{lightgray}{rgb}{0.83, 0.83, 0.83}

\usepackage[colorlinks=true,citecolor=OliveGreen,linkcolor=OliveGreen]{hyperref}

\usepackage{epstopdf}

\usepackage{csquotes}

\usepackage{type1ec} %

\usepackage{colortbl}

\DeclareMathAlphabet\mathbfcal{OMS}{cmsy}{b}{n}

\allowdisplaybreaks

\emergencystretch=\maxdimen
\newcommand\Tstrut{\rule{0pt}{4.5ex}}         % = `top' strut

\begin{document}

%\preprint{APS/123-QED}

\title{Rotational dynamics of a disk in a thin film of weakly nematic fluid \\ subject to linear friction}

\author{Abdallah Daddi-Moussa-Ider}
\email{abdallah.daddi-moussa-ider@open.ac.uk}
\thanks{corresponding author.}

\affiliation{School of Mathematics and Statistics, The Open University, Walton Hall, Milton Keynes MK7 6AA, UK}

\author{Elsen Tjhung}

\affiliation{School of Mathematics and Statistics, The Open University, Walton Hall, Milton Keynes MK7 6AA, UK}

\author{Marc Pradas}

\affiliation{School of Mathematics and Statistics, The Open University, Walton Hall, Milton Keynes MK7 6AA, UK}

\author{Thomas Richter}

\affiliation{Institut f\"ur Analysis und Numerik, Otto-von-Guericke-Universit\"at Magdeburg, Universit\"atsplatz 2, 39106 Magdeburg,
Germany}

\author{Andreas M. Menzel}
\email{a.menzel@ovgu.de}

\affiliation{Institut f\"ur Physik, Otto-von-Guericke-Universit\"at Magdeburg, Universit\"atsplatz 2, 39106 Magdeburg, Germany}

\date{\today}

\begin{abstract}
Dynamics at low Reynolds numbers experiences recent revival in the fields of biophysics and active matter. While in bulk isotropic fluids it is exhaustively studied, this is less so in anisotropic fluids and in confined situations. Here, we combine the latter two by studying the rotation of a disk-like inclusion in a uniaxially {anisotropic}, globally oriented, incompressible two-dimensional fluid film. In terms of a perturbative expansion in parameters that quantify anisotropies in viscosity and in additional linear friction with a supporting substrate or other type of confinement, we derive analytical expressions for the resulting hydrodynamic flow and pressure fields as well as for the resistance and mobility coefficients of the rotating disk. It turns out that, in contrast to translational motion, the solutions remain well-behaved also in the absence of the additional linear friction. Comparison with results from finite-element simulations show very good agreement with those from our analytical calculations. Besides applications to describe technological systems, for instance, in the area of microfluidics and thin cells of aligned nematic liquid crystals, our solutions are important for quantitative theoretical approaches to fluid membranes and thin films in general featuring a preferred direction. 
\end{abstract}

\maketitle

\section{Introduction}

Proteins play an essential role in the structure, function, and regulation of cells and tissues in living organisms~\cite{engel2008structure, sych2022lipid}.
A deep understanding of the molecular mechanisms underlying membrane processes requires detailed knowledge of the movements of membrane proteins. 
Relating the motion of proteins to the functional state of the membrane can offer insights into various bio-cellular processes such as active transport, transmembrane signaling, and electron transport within biological membranes~\cite{rochaix2011regulation, jaqaman2012regulation, albers2012membrane}. 
Studying protein diffusion in membranes can also enhance our understanding of membrane dynamics and organization, which is crucial for developing targeted drug delivery systems and improving the efficacy of membrane-based therapeutic interventions~\cite{marinko2019folding, lo2023application}.

Proteins exhibit motion across a broad spectrum of time and length scales.
Within phospholipid bilayers, there are two primary modes of motion. These are lateral diffusion within the bilayer plane and rotational diffusion around the bilayer normal axis~\cite{cherry1975protein, kinsey1981dynamics, gall1982protein, lewis1985nmr, das2015membrane}.
Measurements of rotational motion are particularly valuable due to their sensitivity to the shape and size of proteins or their assemblies~\cite{thomas1986rotational}.
Numerous methods and experimental techniques are available in scientific research and engineering for measuring their diffusion~\cite{edidin1987rotational, marsh1993progressive, miao2022paramagnetic}. 
Specifically, rotational diffusion can be measured using various optical methods or by means of saturation-transfer electron-spin resonance (ESR).

In the context of biological signal transmission, protein--protein interaction plays a crucial role. Effective communication often requires proteins to physically contact each other and orient themselves correctly to interact with mutual binding sites, a process that relies primarily on their rotational diffusion~\cite{li2009translational}.
Another example is molecular motors embedded in the membrane, with a portion performing external work, such as rotating a flagellum~\cite{minamino2008molecular, armitage2020assembly}. 
This action exerts a countertorque on the membrane, leading to its rotational motion.

Early theoretical studies on the diffusion of particles within membranes were pioneered by Saffman and Delbr\"uck~\cite{saffman1975brownian, saffman1976brownian}. 
They determined the hydrodynamic resistance experienced by a slender disk moving within a thin layer of viscous fluid. This thin fluid film represents the environment of the lipid bilayer in the membrane.
According to their model, the translational and rotational resistance coefficients, denoted as \( R_\mathrm{T} \) and~\(R_\mathrm{R} \), respectively, for a disk-like inclusion of radius \(a\) in the membrane were obtained as~\cite{saffman1975brownian}
\begin{equation}
    R_\mathrm{T} = \frac{4\pi\eta_\mathrm{S}}{ \ln \left( 
\operatorname{Bq} \right) -\gamma} \, , \qquad
    R_\mathrm{R} = 4\pi \eta_\mathrm{S} a^2 \, .
    \label{eq:first-eqn}
\end{equation}
Here, \(\eta_\mathrm{S}\) is the two-dimensional membrane viscosity, and 
{\( \operatorname{Bq} = \eta_\mathrm{S} / \left( \eta a \right)  \)} is a dimensionless quantity known in interfacial rheology as the Boussinesq number~\cite{manikantan2020surfactant}, with $\eta$ denoting the three-dimensional viscosity of a bulk fluid surrounding the thin fluid layer.
In addition, \(\gamma\) {denotes Euler’s} constant.
The diffusion coefficients are then derived from the resistance coefficients using the classical Stokes--Einstein relations~\cite{einstein05, miller1924stokes}.

It is worth noting that the Saffman--Delbrück theory is applicable in the limit of large~\(\operatorname{Bq}\).
The general case of a finite Boussinesq number was later examined~\cite{hughes1981translational, Stone.Ajdari1998}.
Based on hydrodynamic membrane models~\cite{saffman1975brownian, naji2007corrections}, the viscosity of the lipid membrane can be determined by measuring both the rotational and translational diffusion coefficients of membrane-linked particles, along with their radii~\cite{hormel2014measuring}.
{
In these analyses, the membrane is treated as a thin liquid layer of finite thickness.}

{
A simple phenomenological approach has later been introduced by Evans and Sackmann \cite{evans1988translational} to characterize the weak dynamic coupling of a fluid membrane to an adjacent solid substrate. 
Utilizing this framework, the inertialess equations of motion for membrane flow have been solved for the steady translation and rotation of a disk-like particle.
Remarkably, the resulting drag coefficients show functional dependencies on the dimensionless particle size that are very similar to those derived for particle motion in a membrane bounded by semi-infinite fluid domains. Yet, the scaling with the particle size differs~\cite{hughes1981translational}.
}

{
According to the Evans--Sackmann model, the translational and rotational resistance coefficients for a disk-like inclusion of radius \(a\) in the membrane were determined in a scaled form as
\begin{equation}
    \frac{R_\mathrm{T}}{4\pi\eta_\mathrm{S}} = 
    \frac{c^2}{4} + c\, \frac{K_1(c)}{K_0(c)} \, , \quad
    \frac{R_\mathrm{R}}{4\pi a^2 \eta_\mathrm{S}} = 
    1 + \frac{c}{2} \frac{K_0(c)}{K_1(c)}  \, , 
    \label{eq:first-eqn-evans-sackmann}
\end{equation}
where $c$ is a dimensionless parameter defined as $c= a\left(b_\mathrm{S} / \eta_\mathrm{S} \right)^{1/2}$ and $b_\mathrm{S}$ marks an intrinsic coefficient of friction between the membrane and the substrate.
Expanding the right-hand sides of Eqs.~\eqref{eq:first-eqn-evans-sackmann} in a power series of \( c \) up to order $c^2$ yields Eq.~\eqref{eq:first-eqn}, which is derived using the Saffman--Delbrück model, for \( c = 2/\operatorname{Bq} \).
Our contribution relates to the Evans--Sackmann model due to its simplicity and practicality, when compared to the original Saffman--Delbrück model, particularly in the presence of a substrate.}

Confining boundaries and dampening interfaces are known to reduce the {mobility} coefficient of microparticles, leading to decreased diffusion
\cite{han09, mendoza2014particle, hosaka2021hydrodynamic}.
Along these lines, effects of the system size on the rotational diffusion of membrane proteins have been explored using molecular dynamics simulations~\cite{voegele2019finite, voegele2016divergent, marrink2019computational}.
Observations include a decreased rotational diffusion coefficient when compared to an infinitely {extended} bulk system, influenced by hydrodynamic interactions with the confining boundaries~\cite{voegele2019finite}.
Specifically, the rotational dynamics of a sample protein under cylindrical nanopore confinement has been examined~\cite{haridasan2019rotational}. 
The simulation study revealed a twofold reduction in the rotational diffusion coefficient compared to the bulk value when the confinement radius is twice the hydrodynamic radius of the protein.
A versatile numerical scheme for predicting diffusion coefficients of arbitrarily shaped objects embedded in lipid bilayer membranes has also been presented~\cite{camley2013diffusion}.

The rotational dynamics of membrane proteins have been investigated using molecular-dynamics simulations of coarse-grained membranes and two-dimensional Lennard-Jones fluids under different crowding conditions~\cite{javanainen2020rotational}.
It has been observed that the Saffman--Delbr\"uck theory~\cite{saffman1975brownian} accurately describes the size-dependence of rotational diffusion under dilute conditions, whereas more complex scaling laws emerge under crowded conditions.
{The study reports a crowding-induced transition between Saffman--Delbrück-like and stronger scanlings, where the rotational diffusion coefficient \(D_\mathrm{R} \sim a^{-\nu}\) with \(\nu > 2\). In systems of increasing levels of crowding, the Saffman--Delbrück model fits the data for small disks, while larger disks exhibit a stronger scaling of $\nu=3$. 
}
More recently, the anisotropic diffusion of membrane proteins has been computationally investigated~\cite{javanainen2021anisotropic}. 
Generally, anisotropy can arise from the anisotropic shape of proteins and protein complexes or from the anisotropic interactions of proteins with the lipid bilayer~\cite{han2006brownian, dlugosz2014transient, ellena2011membrane}.

In mixtures containing insoluble surfactants, phase separation commonly occurs, often resulting in the formation of two-dimensional dispersions. These dispersions consist of densely packed, ordered aggregates floating within a disordered fluid phase~\cite{lingwood2010lipid, sezgin2017mystery}. 
Inspired by such and other investigations, we examined in our previous contribution~\cite{daddi2024hydrodynamics} the translational dynamics of inclusions within a thin, two-dimensional fluid film. Yet, in that work, we moved one step beyond the previous considerations by asking the question of the influence of possible orientational order of the fluid membrane. Specifically, we asked for the translational dynamics of a circular disk laterally surrounded under no-slip conditions by a globally orientationally ordered, uniaxially anisotropic, incompressible two-dimensional fluid layer. 

In this context, nematic liquid crystals come to our mind as a fluid example system. Generally, they exhibit phases of matter that possess physical characteristics intermediate between those of disordered isotropic liquids and ordered crystalline solids~\cite{de1993physics}. 
Particularly, when they consist of elongated rod-like (or flat disk-like) organic molecules they are well-known candidates of forming the mentioned uniaxially anisotropic phases. 
The local axis of collective orientational ordering is termed \enquote{director}~\cite{lavrentovich2016active}.

Beyond the focus of our previous work on translational motion~\cite{daddi2024hydrodynamics}, we now address rotational motion of the circular disk in a globally uniaxially ordered, incompressible two-dimensional fluid at low Reynolds numbers. As before, we include the influence of additional linear friction, which may be induced by a supporting substrate or other types of environment of the two-dimensional fluid film or membrane. 
{We note in this context that in the previous translational case the additional linear friction has a stabilizing effect for the system and the associated mathematical solution. In an infinitely extended two-dimensional fluid, the mathematical solution of the Stokes equation for the flow resulting from the net translational motion of an object immersed under no-slip conditions diverges logarithmically as a function of the distance from that object. This is referred to as the Stokes paradox. Additional linear friction stabilizes the solution and cancels the divergence. Yet, such an effect is absent in the case of rotational motion, so that additional linear friction does not take the central role of stabilization. Generally for the flow field, the leading order of dependence on the distance from the object for its rotational motion is obtained by derivation of the leading order for its translational motion \cite{dhont1996introduction, kim2013microhydrodynamics, puljiz2019displacement, richter22}. Therefore, we find decay with inverse distance instead of logarithmic divergence in the rotational case.}

More in detail, %To this end, 
we here utilize a technique of two-dimensional Fourier transformation to solve the hydrodynamic equations for the velocity and pressure fields~\cite{bickel06, felderhof06, daddi17, daddi16c, daddi18epje, daddi18jcp}.
In cases of weak anisotropy of both viscosity and of linear friction, we apply a perturbative approach up to second order in these small parameters. We derive approximate expressions for the hydrodynamic fields, resistance and mobility coefficients. Specifically and in contrast to the translational case \cite{daddi2024hydrodynamics}, in the absence of additional linear friction we obtain exact analytical expressions in the form of convergent series expansions.
Our theoretical predictions are supplemented by a comparison with results from finite-element simulations, showing very good agreement between the two different approaches.

\section{Flow in two-dimensional anisotropic fluids with friction}

The equations describing the dynamics of low-Reynolds-number flow in a two-dimensional anisotropic fluid medium are derived in our previous publication~\cite{daddi2024hydrodynamics}.
In the following, we summarize the relevant equations for the further analysis in the present framework.

Throughout, we consider the regime of low Reynolds numbers, where viscous forces dominate over inertial forces. In such scenarios, the dynamics of the fluid are primarily governed by the linear Stokes equation. 
Here, we investigate a generalized version of the Stokes equations to incorporate spatially homogeneous uniaxial anisotropy of the fluid under consideration and potential frictional effects from the surrounding environment. In practice, this friction may stem, for instance, from a substrate supporting a two-dimensional thin fluid film or a slab-type confinement of a fluid layer between two parallel plates widely extended plates.

In such a configuration, the general expression for the flow equation can be formulated as
\begin{equation}
    \boldsymbol{\nabla} \cdot \boldsymbol{\sigma} (\bm{r})
     + \bm{M} \cdot \bm{v}(\bm{r}) = \bm{f} (\bm{r})  \, , 
\end{equation}
wherein 
$ \boldsymbol{\sigma} (\bm{r}) = p(\bm{r}) \, \bm{I} + \widetilde{\boldsymbol{\sigma}} (\bm{r}) $
denotes the stress tensor. Here, $p(\bm{r})$ represents the pressure field, 
$\bm{I}$~stands for the unit tensor, and 
$\widetilde{\boldsymbol{\sigma}} (\bm{r})$ incorporates the effect of viscous dissipation into the dynamics. Moreover, 
$\bm{f} (\bm{r})$ denotes the two-dimensional inplane force density acting on the fluid 
and 
$\bm{M}$
represents the tensor of linear friction associated with the bottom and/or top environment. $\bm{v} (\bm{r})$ is the velocity field.

In this contribution, we investigate a uniaxially symmetric, anisotropic fluid medium characterized by the local orientation of the axis of uniaxial order. This orientation is described by the normalized director field~$\bm{\hat{n}}(\bm{r})$. In the present work, we explore a scenario of the director being consistently and uniformly aligned throughout the system along one persistent single global axis, as considered in previous studies as well~\cite{kneppe81,heuer92, daddi2018dynamics}. 
In reality, such an alignment can be achieved, for example, through the influence of a strong aligning external electric or magnetic field~\cite{degennes95,pleiner96,pleiner02} or a corresponding preparation of the supporting substrate and/or confining plates.
Without sacrificing generality, we select $\bm{\hat{n}}$ to be aligned along the $\bm{\hat{x}}$ direction of our Cartesian coordinate frame.

{
In this situation, the stress tensor \(\bm{\widetilde{\sigma}} (\bm{r})\) is reduced to account only for the pure dissipative stress arising from gradients in the fluid flow~\cite{pleiner96,pleiner02},
\begin{equation}
\widetilde{\sigma}_{ij} = \sigma_{ij}^{\text{D}} = {}-\nu_{ijkl}\nabla_l v_k \, . \label{eq:momentumEquation}
\end{equation}
In this expression, $\nu_{ijkl}$ refer to the components %of the surface viscosity tensor of uniaxial symmetry of SI unit kg/s, given by \cite{pleiner96}
of the viscosity tensor of uniaxial symmetry \cite{pleiner96},
\begin{align}
         \nu_{ijkl} &= \nu_2(\delta_{ik}\delta_{jl} + \delta_{il}\delta_{jk})
+ \bar{\nu} n_in_jn_kn_l + (\nu_4-\nu_2)\delta_{ij}\delta_{kl} \notag \\[3pt]
&\quad+ (\nu_3-\nu_2)(n_in_k\delta_{jl}+n_in_l\delta_{jk}+n_jn_k\delta_{il}+n_jn_l\delta_{ik}) \notag \\[3pt]
&\quad+ (\nu_5-\nu_4+\nu_2)(\delta_{ij}n_kn_l+\delta_{kl}n_in_j) \, , \label{eq:nu_ijkl}
\end{align}
where $\nu_i$ ($i \in 1, \dots, 5$) are five coefficients of 
viscosity for the case of uniaxial symmetry~\cite{pleiner96}. % of SI unit kg/s in two spatial dimensions.
Here, we introduced the abbreviation 
\begin{equation}
    \bar{\nu} = 2 \left( \nu_1 + \nu_2 - 2\nu_3 \right) \, .
\end{equation}
%For an isotropic fluid medium, it follows that for \(i = 1, \dots, 4\), \(\nu_i = \eta_\mathrm{S}\), defined in Eqs.~\eqref{eq:first-eqn-evans-sackmann}, and \(\nu_5 = 0\).}
In the case of an isotropic fluid, the first term on the right-hand side of Eq.~(\ref{eq:nu_ijkl}) of coefficient $\nu_2$ refers to the shear viscosity and thus here is related to the viscosity $\eta_\mathrm{S}$. The third term of coefficient $\nu_4-\nu_2$ relates to volume/bulk (or area in two dimensions) viscosity. All other terms are absent for an isotropic fluid as they explicitly involve the role of the director $\bm{\hat{n}}$.} 
We remark that the convention for the sign of $\widetilde{ \boldsymbol{\sigma} }(\bm{r})$ can vary across different sources in the literature and, accordingly, the associated tensor might be defined with a minus sign in some cases.

As typically assumed in corresponding analyses of fluid flows, we prescribe local volume conservation and a constant fluid density. Consequently, the continuity equation can be expressed as
\begin{equation}
    \boldsymbol{\nabla} \cdot \bm{v}(\bm{r})=0 \, . \label{eq:IncompressGleichung}
\end{equation}
Moreover, in the following discussion, we operate under the assumption that the friction tensor is symmetric, taking the form
\begin{equation}
    \bm{M} = 
    \begin{pmatrix}
        m_\parallel^2 & 0 \\
        0 & m_\perp^2 
    \end{pmatrix} .
\end{equation}

As noted above, the equations governing the dynamics of a two-dimensional weakly nematic fluid subject to linear friction have been derived in our previous  publication~\cite{daddi2024hydrodynamics}. They can be projected into Cartesian coordinates as
\begin{subequations}
\begin{align}
    p_{,x} -\nu_3 \left( \lambda_+ v_{x,xx} + v_{x,yy} - \alpha_\parallel^2 v_x \right)  &= f_x \, , \\[3pt]
    p_{,y} -\nu_3 \left( v_{y,xx} + \lambda_- v_{y,yy} - \alpha_\perp^2 v_y \right) &= f_y \, ,
\end{align}
\label{eq:gov_eqns_fourier_final}
\end{subequations}
where we define the dimensionless viscosities 
\begin{equation}
    \lambda_\pm = 1 + \frac{A}{2} \pm b \, .
    \label{eq:lambda_pm}
\end{equation}
The parameters
    \begin{align}
    A &= 2\left( \frac{\nu_1+\nu_2}{\nu_3} - 2 \right) , \label{eq:A} \\[1pt]
    b &= \frac{\nu_1-\nu_4+\nu_5}{\nu_3} \label{eq:b}
\end{align}
represent two dimensionless numbers related to anisotropy.
Additionally, for convenience, we introduce the abbreviations  
$\alpha_\parallel^2 = m_\parallel^2 / \nu_3$ and $\alpha_\perp^2 = m_\perp^2 / \nu_3$, respectively. We note that both $\alpha_\parallel^2$ and $\alpha_\perp^2$ possess dimensions of inverse length squared.

To derive the solution for the hydrodynamic field, we utilize a two-dimensional Fourier transformation in both the $x$- and $y$-directions. For any scalar or vector function $g(\boldsymbol{\rho}) = g(x,y)$, where $\boldsymbol{\rho} = (x,y)$, its two-dimensional Fourier transformation is denoted as $\widetilde{g} (\bm{k})$, where $\bm{k}$ represents the wavevector. We introduce the wavenumber $k = |\bm{k}|$, along with the unit wavevector $\hat{ \bm{k} } = \bm{k}/k$. Furthermore, we define a system of polar coordinates such that $\hat{ \bm{k} } = (\cos\phi, \sin\phi)$.

The solution for the velocity field in Fourier space can be obtained by taking the Fourier transformations of Eqs.~\eqref{eq:gov_eqns_fourier_final} and then applying the divergence-free condition stated by Eq.~\eqref{eq:IncompressGleichung} to eliminate the pressure field to obtain
\begin{equation}
    \widetilde{\bm{v}} (k, \phi) = 
    \widetilde{\bm{\mathcal{G}}} (k, \phi) \cdot \widetilde{\bm{f}} (k, \phi) \, ,
    \label{eq:GeschwindFourier}
\end{equation}
wherein the Green's function is denoted as
\begin{equation}
    \widetilde{\bm{\mathcal{G}}} (k, \phi) =
    \left( \nu_3 H (k, \phi) \right)^{-1}
    \bm{G} (\phi) \, . 
\end{equation}
Here,
\begin{equation}
    H (k, \phi) = \alpha_\parallel^2 \sin^2\phi + \alpha_\perp^2 \cos^2\phi + k^2 B(\phi)
\end{equation}
with the abbreviation
\begin{equation}
    B(\phi) = 1 + A \sin^2\phi \cos^2\phi \, .
\end{equation}
In addition, 
\begin{equation}
    \bm{G} (\phi) = 
    \begin{pmatrix}
        \sin^2 \phi        & -\sin\phi \cos\phi \\
        -\sin\phi \cos\phi & \cos^2\phi
    \end{pmatrix} . 
\end{equation}

Our analysis is limited to the situation of $A \geq -4$, ensuring that $B(\phi) \geq 0$. This requirement is crucial to guarantee $H(k,\phi) > 0$, which is necessary for the Green's function to remain well-defined across all values of $k$.
The condition is linked to thermodynamic stability, necessitating that $A > \nu_2/\nu_3 - 4$. Since both $\nu_2$ and $\nu_3$ must be positive, we thus require $A > -4$.
The solution in real space can then be obtained via inverse Fourier transformation.

\section{Hydrodynamic flow fields}

\begin{figure}
    \centering
    \includegraphics[scale=.8]{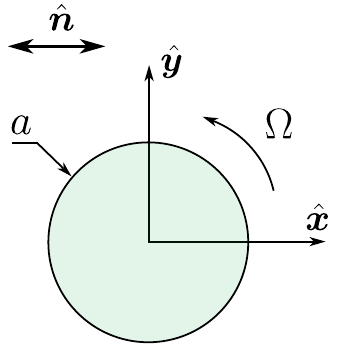}
    \caption{
    Schematic illustration of the quantities involved in the rotational motion of the thin disk under consideration in a surrounding two-dimensional sheet of a uniaxially anisotropic fluid. The circular disk of radius \(a\) rotates at low-Reynolds-number conditions with an angular velocity \(\Omega\). Considering global homogeneous orientational alignment of the uniaxial fluid, the axis of anisotropy is characterized by the nematic director \(\hat{\bm{n}}\). It is consistently aligned along the \(x\)-direction.
    }
    \label{fig:disk}
\end{figure}

We investigate the steady rotational motion of a circular disk within a two-dimensional anisotropic sheet of uniaxial nematic fluid.
In our analysis, we designate $a$ as the radius of the disk and $\Omega$ as its angular velocity, see Fig.~\ref{fig:disk}.

First, our focus is on analyzing the hydrodynamic velocity field surrounding the disk.
To this end, we represent the unknown force distribution that the disk applies to the fluid along its boundary using a Fourier series
\begin{equation}
    \bm{f} (\rho, \theta) = \delta(\rho-a) \sum_{n=-\infty}^\infty \bm{f}_n  \, e^{in\theta} \, . 
    \label{eq:Kraft-distr}
\end{equation}
The coefficients $\bm{f}_n$ are determined below based on the specified boundary condition.
We express the force density  
in Fourier space as
\begin{equation}
    \widetilde{\bm{f}} (k, \phi) = 
    2\pi a \sum_{n=-\infty}^\infty i^{-n} \, \bm{f}_n J_n(ka) \, e^{in\phi} \, .
    \label{eq:Kraft_fourier}
\end{equation}
$J_n$ represent the Bessel functions of the first kind of $n$th degree~\cite{abramowitz72}.

Our primary objective is to determine the force density distribution along the circumference of the disk that meets the boundary conditions for the velocity field.
We apply no-slip boundary conditions along the circumference. Specifically, for the surrounding fluid this condition is expressed  as $\bm{v}(\rho=a) = a \Omega \, \hat{\boldsymbol{\theta}}$. Here, $\hat{\boldsymbol{\theta}} = -\sin\theta \, \hat{\bm{x}} + \cos\theta \, \hat{\bm{y}}$ represents the unit azimuthal vector in cylindrical polar coordinates.
The 
Cartesian components of the velocity field at the circumference of the disk are therefore prescribed as $v_x(\rho=a,\theta) = -a\Omega \sin\theta$ and $v_y(\rho=a,\theta) = a\Omega \cos\theta$. 
Matching the Fourier modes at the boundary of the disk reveals that only the odd Fourier modes contribute to the force density, see below.

Applying the inverse Fourier transformation, the velocity field is expressed in real space in the form~\cite{baddour2011two}
\begin{equation}
    \bm{v}(\rho, \theta) = 
    \left( 2\pi \right)^{-2} 
    \sum_{m=-\infty}^\infty i^m \mathbfcal{S}_{m} (\rho) \, e^{im\theta} \, , \label{eq:v_inv_fourier}
\end{equation}
wherein~$\mathbfcal{S}_{m}$ is a function of $\rho$ defined as the double integral over $k$ and $\phi$,
\begin{equation}
    \mathbfcal{S}_{m} (\rho) = 
    \int_0^{2\pi} \int_0^\infty \widetilde{\bm{v}} (k,\phi) \,
     J_{m} (k\rho) \, e^{-im\phi} \, k \, \mathrm{d}k \, \mathrm{d}\phi \, .
     \label{eq:Sm}
\end{equation}
The boundary condition prescribed at the circumference of the disk corresponds to a constant angular velocity. Therefore, it follows from  Eq.~\eqref{eq:v_inv_fourier} that the correct velocity at \(\rho = a\) is obtained for
\begin{subequations}\label{eq:Sm_rho}
    \begin{align}
    \mathbfcal{S}_{\pm m}(\rho = a) \cdot \hat{\bm{x}}
    &= \phantom{\pm} \frac{ \delta_{|m| 1} }{2 } \, 
    \left( 2\pi \right)^{2} a \Omega \, ,  \\
    \mathbfcal{S}_{\pm m}(\rho = a) \cdot \hat{\bm{y}}
    &= \pm\frac{ \delta_{|m| 1} }{2i} \, 
    \left( 2\pi \right)^{2} a \Omega \,  .
\end{align} 
    \label{eq:BC}
\end{subequations}
Thus, only the first mode matches the prescribed velocity of the disk at \(\rho = a\), while all other modes are zero.

Then, by utilizing Eqs.~\eqref{eq:GeschwindFourier} and \eqref{eq:Kraft_fourier}, we find that the integration with respect to \(\phi\) in \(\mathbfcal{S}_m\), as defined by Eq.~\eqref{eq:Sm}, vanishes when \(m-n\) is an odd number. Given that Eqs.~\eqref{eq:Sm_rho} represent a system of linear equations for the unknown values of \(\bm{f}_n\), it follows that all even Fourier modes in the expression of the force density must vanish. Consequently, only the odd modes will persist in the series expansion of the velocity field.

Subsequently, by accounting only for the odd Fourier modes, the velocity field can be expressed as an infinite double sum of the form,
\begin{equation}
    \bm{v}(\rho, \theta) = \frac{a}{2\pi\nu_3} 
    \sum_{m=-\infty}^\infty e^{i(2m+1)\theta} \sum_{n = -\infty}^\infty \, \mathbfcal{V}_{mn} (\rho) \, , 
\end{equation}
where
\begin{equation}
    \mathbfcal{V}_{mn} (\rho) = \int_0^{2\pi}
    \Gamma_{mn} (\rho, \phi) \, e^{2i (n-m) \phi}  \, \bm{G} (\phi) \cdot \bm{f}_{2n+1} \, \mathrm{d} \phi \, 
    \label{eq:Q}
\end{equation}
and 
\begin{equation}
    \Gamma_{mn} = \frac{(-1)^{m+n}}{B}  \int_0^\infty \frac{u \, J_{2m+1}(u) J_{2n+1}({w} u)}{u^2 + \sigma^2} \,  \, \mathrm{d} u \, . 
    \label{eq:GAMMA}
\end{equation}
In these expressions, we have defined the dimensionless numbers ${w} = a/\rho$ and $\sigma$ according to
\begin{equation}
    \sigma^2 = \frac{\rho^2}{B}
    \left( \alpha_\parallel^2 \sin^2\phi + \alpha_\perp^2 \cos^2\phi \right) .
\end{equation}
Here, we have introduced the change of variables $u = k\rho$. The condition ${w} \le 1$ denotes the fluid region outside the disk. We recall that the index~$m$ is used for the inverse Fourier transformation of the velocity field, while the index~$n$ applies to the Fourier representation of the unknown force density.

Next, we set $\bm{f}_{2n+1} = \left( ic_{2n+1}, -d_{2n+1} \right)^{{\top}} $ with \((\cdot)^\top\) denoting the transpose.
These coefficients are chosen as real numbers. Their values are determined based on the boundary conditions prescribed at the surface of the disk, as detailed below.
To ensure the coefficients $c_{2n+1}$ and~$d_{2n+1}$ are real-valued, we additionally require that $c_{-(2n+1)} = -c_{2n+1}$ and $d_{-(2n+1)} = d_{2n+1}$.
Therefore, the force density given by Eq.~\eqref{eq:Kraft-distr} can be expressed as
\begin{equation}
    \bm{f} = 2 \delta(\rho-a)
    \sum_{n=0}^\infty 
    \begin{pmatrix}
        -c_{2n+1} \sin \left( (2n+1)\theta \right) \\[3pt]
        \,\, d_{2n+1} \cos \left( (2n+1)\theta \right) \,\,
    \end{pmatrix} . \label{eq:Kraft-distr-simplified}
\end{equation}
The corresponding Fourier representation is given by Eq.~\eqref{eq:Kraft_fourier}. It can be expressed as
\begin{equation}
    \widetilde{\bm{f}} = 
    \sum_{n=0}^\infty s_{2n+1} 
    \begin{pmatrix}
        c_{2n+1} \sin \left( (2n+1)\phi \right) \\[3pt]
        \,\, d_{2n+1} \cos \left( (2n+1)\phi \right) \,\,
    \end{pmatrix} 
\end{equation}
with 
\begin{equation}
    s_{2n+1} = 4i\pi a (-1)^n J_{2n+1} (ka) \, .
\end{equation}

Consequently, the velocity field is given by
\begin{align}
    \bm{v} (\rho, \theta) &=  
    \frac{a}{\nu_3} 
    \sum_{m = 0}^\infty \sum_{n = 0}^\infty
    \left(  U_{mn} \, \bm{\hat{x}} 
    + V_{mn}  \, \bm{\hat{y}}  \right) , 
\end{align}
wherein
\begin{subequations} 
    \begin{align}
    U_{mn} &= - \frac{2}{\pi} \int_0^{2\pi} 
    \mathcal{M}_{mn} (\theta, \phi) \, \Gamma_{mn} (\rho, \phi) \, \sin \phi \, \mathrm{d}\phi \, , \\[3pt]
    V_{mn} &= \phantom{+} \frac{2}{\pi} \int_0^{2\pi} 
    \mathcal{M}_{mn} (\theta, \phi) \, \Gamma_{mn} (\rho, \phi) \, \cos \phi \, \mathrm{d}\phi \, ,
\end{align} \label{eq:UV}
\end{subequations}
while
\begin{equation}
    \mathcal{M}_{mn} (\theta, \phi) = 
    \cos \left( (2m+1)(\phi-\theta) \right)
    Z_{2n+1} (\phi) \, ,
\end{equation}
with 
\begin{align}
    Z_{2n+1} (\phi) &= c_{2n+1} \sin \left( (2n+1)\phi \right) \sin\phi \notag \\
    &\quad- d_{2n+1} \cos \left( (2n+1)\phi \right) \cos\phi \, .
\end{align}

As can be inferred from Eqs.~\eqref{eq:GAMMA} and \eqref{eq:UV}, determining the velocity field involves evaluating a double integral. Given the complexity of an analytical evaluation, particularly in the general case, we adopt the following approach.

The integral over the scaled wavenumber, as defined by Eq.~\eqref{eq:GAMMA}, allows for exact evaluation, resulting in an analytical expression. However, for the subsequent integrals over \(\phi\) given in Eqs.~\eqref{eq:UV}, we  resort to approximations.

By employing complex calculus, it can be demonstrated that $\Gamma_{mn}$, as defined by Eq.~\eqref{eq:GAMMA}, can be accurately calculated as,
\begin{equation}
    \Gamma_{mn} = \frac{1}{B} 
    \begin{cases}
        \Delta_{mn} & \text{if } m \le n \, , \\
        \Delta_{mn} - \mathcal{P} \Delta_{mn} & \text{if } m > n \, ,
    \end{cases} \label{eq:improper_int}
\end{equation}
where 
\begin{equation}
\label{eq:Delta}
    \Delta_{mn} = I_{2n+1}({w}\sigma) \, K_{2m+1}(\sigma) \, , 
\end{equation}
with \( I_{2n+1} \) and \( K_{2n+1} \) denoting the $(2n+1)$th-order modified Bessel functions of the first and second kind, respectively.
Here, \(\mathcal{P} \Delta_{mn}\) denotes the principal part of the series expansion of \(\Delta_{mn}\) around \(\sigma = 0\). 
Specifically, the principal part corresponds to the part of the Laurent series expansion of \(\Delta_{mn}\) characterized by negative exponents~\cite{carrier2005functions}. This result is valid only when \(w \le 1\), which corresponds to the physical scenario of the fluid area outside the disk. A detailed analytical solution of a similar integral is provided in Appendix~B of Ref.~\onlinecite{daddi2024hydrodynamics}.

In Table \ref{tab:PP}, we present the corresponding values of $\mathcal{P} \Delta_{n+1,n}$, $\mathcal{P} \Delta_{n+2,n}$, and $\mathcal{P} \Delta_{n+3,n}$ for $n = 0, \dots, 4$.
To obtain expressions for $\mathcal{P} \Delta_{mn}$ where $m>n$, one can derive them for any given values of $m$ and $n$ using computer algebra systems
\cite{maple24}. 
This can be accomplished by computing the Laurent series expansion of $\Delta_{mn}$ and considering only the terms with negative exponents.

\begin{table*}
		\centering
		{\renewcommand{\arraystretch}{2.3}
			\begin{tabular}{|c|c|c|c|}
				\hline
				~$n$~ & $m=n+1$ & $m=n+2$ & $m=n+3$ \\
				\hline
				0 & $4wt$ & $-12wt \left( 1-2w^2-16t\right)$ & $8wt \left[ 3-15w^2+15w^4-120 \left( 1-3w^2\right)t + 2880 t^2 \right]$ \\
				\arrayrulecolor{lightgray}\hline
				1 & $8w^3 t$ & $-20w^3 t \left( 2-3w^2-48t \right)$ & $12w^3 t \left[ 10-35w^2+28w^4-280 \left( 2-4w^2\right)t + 17920 t^2 \right]$ \\
				\hline
				2 & $12w^5 t$ & $-28w^5t \left( 3-4w^2-96t \right)$ & $16w^5 t \left[ 21-63w^2+45w^4 - 504 \left( 3-5w^2\right)t + 60480 t^2 \right]$ \\
				\hline
				3 & $16w^7 t$ & $-36w^7t \left( 4-5w^2-160t \right)$ & $20w^7 t \left[ 36-99w^2+66w^4-792 \left( 4-6w^2\right)t + 152064 t^2 \right]$ \\
				\hline
				4 & $20w^9t$ & ~~$-44w^9t \left( 5-6w^2-240t \right)$~~ & ~~$24w^9 t \left[ 55-143w^2+91w^4 - 1144 \left( 5-7w^2 \right)t + 320320 t^2 \right]$~~ \\
				\hline
				\vdots & \vdots & \vdots & \vdots \\
				\hline
				$n$ & $\mathcal{P} \Delta_{n+1, n}$ & $\mathcal{P} \Delta_{n+2, n}$ & $\mathcal{P} \Delta_{n+3, n}$ \\
				\arrayrulecolor{black}\hline
			\end{tabular}
		}
		\caption{
		The principal part of the series expansion of $\Delta_{mn}$ for cases where $m=n+1$, $m=n+2$, and $m=n+3$, as defined by Eq.~\eqref{eq:Delta}. General mathematical expressions applicable to various values of $n$ are given by Eqs.~\eqref{eq:PP3}. Here, we define $t=1/\sigma^2$.
		}
		\label{tab:PP}
\end{table*}

By defining $t=1/\sigma^2$, we obtain the initial terms of $\mathcal{P} \Delta_{mn}$ for $m=n+1$, $m=n+2$, and $m=n+3$ as
\begin{widetext}
\begin{subequations} \label{eq:PP3}
	\begin{align}
		\mathcal{P} \Delta_{n+1, n} &= 4(n+1) {w}^{2n+1} t \, , \\
            \mathcal{P} \Delta_{n+2, n} &= -4(2n+3) w^{2n+1} t
            \left[ n+1 - (n+2)w^2 - 8(n+1)(n+2)t \right] , \\
            \mathcal{P} \Delta_{n+3, n} &= 
            4(n+2) w^{2n+1} t
            \big[ (n+1)(2n+3) - (2n+3)(2n+5)w^2 + (n+3)(2n+5)w^4 \notag \\
            &\quad+ 8(2n+3)(2n+5) \left( (n+3)w^2 - (n+1) + 8(n+1)(n+3)t \right) t
            \big] .
        \end{align}
\end{subequations}
\end{widetext}

To quantify the degree of anisotropy due to friction, we define the dimensionless number
\begin{equation}
    \beta = 1 - \frac{\alpha_\perp}{\alpha_\parallel} \, . \label{eq:def_beta}
\end{equation}

\begin{table}
    \centering
    {\renewcommand{\arraystretch}{1.75}
    \begin{tabular}{|c|c|c|c|c|c|}
    \hline
    ~~$n$th order harmonic~~ & ~~$A$~~ & ~~$\beta$~~ & ~~$A^2$~~ & ~~$A\beta$~~ & ~~$\beta^2$~~ \\
    \hline
    0 & $\checkmark$ & $\checkmark$ & $\checkmark$ & $\checkmark$ & $\checkmark$ \\ 
    2 &           & $\checkmark$ &           & $\checkmark$ & $\checkmark$ \\
    4 & $\checkmark$ &           & $\checkmark$ & $\checkmark$ & $\checkmark$ \\
    6 &           &           &           & $\checkmark$ &  \\
    8 &           &           & $\checkmark$ &           &           \\
    \hline
\end{tabular}
}
    \caption{The harmonic contributions (Fourier modes) for each order in the multi-Taylor expansion in the small parameters $A$ and $\beta$. 
    Radial and azimuthal velocities exhibit sine and cosine harmonics, respectively. Notably, the zeroth harmonic holds significance solely in the azimuthal velocity component. It vanishes for the radial velocity. Checks mark the only non-vanishing modes.}
    \label{tab:harmonics_velocity}
%\end{table}
\vspace{0.5cm}
%\begin{table}
    \centering
    {\renewcommand{\arraystretch}{1.75}
    \begin{tabular}{|c|c|c|c|c|c|}
    \hline
    ~~$n$th order harmonic~~ & ~~$A$~~ & ~~$\beta$~~ & ~~$A^2$~~ & ~~$A\beta$~~ & ~~$\beta^2$~~ \\
    \hline
    2 & $\checkmark$ & $\checkmark$ & $\checkmark$ & $\checkmark$ & $\checkmark$ \\
    4 & $\checkmark$ & $\checkmark$ & $\checkmark$ & $\checkmark$ & $\checkmark$ \\
    6 & $\checkmark$ &           & $\checkmark$ & $\checkmark$ & $\checkmark$ \\
    8 &           &           & $\checkmark$ & $\checkmark$ &           \\
    10 &          &           & $\checkmark$ &           &           \\   
    \hline
\end{tabular}
}
    \caption{The sine harmonic contributions (Fourier modes) for each order in the multi-Taylor expansion in the small parameters $A$ and $\beta$ of the pressure. Again, the only non-vanishing modes are indicated by checks. 
    }
    \label{tab:harmonics_pressure}
\end{table}

As previously mentioned, evaluating the integrals in Eqs.~\eqref{eq:UV} is delicate and non-trivial. To simplify the analytical process, we examine the case of weak fluid anisotropy, specifically when $|\beta| \ll 1$. Additionally, we limit our analysis to cases where $|A| \ll 1$ (see Eq.~\eqref{eq:A} for the definition of $A$). Both $A$ and $\beta$ can have positive or negative sign. 

Given these small parameters, the integrands in Eqs.~\eqref{eq:UV} can be expanded as a multi-Taylor series in~$A$ and~$\beta$. This approach allows for a straightforward analytical evaluation of the resulting integrals with respect to $\phi$ in Eqs.~\eqref{eq:UV}, which would otherwise be challenging. Our analysis includes terms up to the second order in the multi-Taylor series expansion.
To determine the unknown series coefficients \(c_{2n+1}\) and \(d_{2n+1}\), we apply the boundary conditions Eqs.~\eqref{eq:BC} for the velocity field at the circumference of the disk. This results in a linear system of equations that we solve for the unknown coefficients of force density.

We express the solution for the hydrodynamic velocity and pressure fields in the form of a Fourier series
\begin{subequations} \label{eq:v_p_form_Fourier}
    \begin{align}
    v_\rho &= a\Omega \, \sum_{n=1}^\infty g_n (\rho) \sin (2n \theta) \, , \\
    v_\theta &= a\Omega \, \sum_{n=0}^\infty h_n (\rho) \cos (2n \theta) \, , \\
    p &= \frac{\Omega \nu_3}{a} \, \sum_{n=1}^\infty u_n (\rho) \sin (2n \theta) \, . \label{eq:p_form_Fourier}
\end{align} 
\end{subequations}

At leading orders in~\(A\) and~\(\beta\), we find that only the second and fourth Fourier modes in the radial and azimuthal velocities, along with the zeroth mode of azimuthal velocity, remain non-vanishing. Additionally, the sixth Fourier mode for pressure persists.
Notably, the zeroth-order mode of the pressure vanishes.
When accounting for second-order terms, the sixth and eighth Fourier modes for the velocity as well as the tenth Fourier mode for the pressure remain non-vanishing. 
We summarize the non-vanishing Fourier modes of the velocity {expressed in terms of \( g_n \) and \( h_n \) as given in the form of} a multi-Taylor expansion in parameters \(A\) and \(\beta\) in Tab.~\ref{tab:harmonics_velocity}, while we present the corresponding results for the pressure {expressed in terms of \( u_n \)} in Tab.~\ref{tab:harmonics_pressure}.

The non-vanishing Fourier modes for the velocity and pressure fields are obtained up to second order in \(A\) as presented in Tab.~\ref{tab:solution-no-firction}.
We infer that the coefficients $u_n$ defining the Fourier series for the pressure field vanish for odd $n$ in the case of \(b=0\).

To enhance clarity,  
we introduce the dimensionless quantities \( r = \alpha_\parallel \rho \) and \( c = \alpha_\parallel a \). They represent the rescaled radial distance and the rescaled radius of the disk, respectively. 
The final expressions for the velocity and pressure fields of the fluid are then written as
\begin{subequations}
    \begin{align}
    \bm{v} &= a \Omega \left( \bm{v}^{\mathrm{IES}} + \bm{v}^* \right) \, , \\
    p &= \frac{\Omega \nu_3}{a} \left( p^{\mathrm{IES}} + p^* \right) \, , 
\end{align}
\end{subequations}
wherein \(\bm{v}^{\mathrm{IES}}\) is the solution {for the {isotropic Evans--Sackmann} model implying} \(A = \beta=b=0\). Specifically, \(\bm{v}^*\) denotes the leading-order correction due to uniaxial anisotropy in viscosity and friction.
{The solution} for the velocity and pressure field {for the isotropic Evans--Sackmann model} is obtained as
\begin{equation}
    v_\rho^{\mathrm{IES}} = 0 \, , \qquad
    v_\theta^{\mathrm{IES}} = { \frac{ K_1(r) }{ K_1(c) } } \, , 
    \qquad
    p^{\mathrm{IES}} = 0 \, .
\end{equation}
% We remark that $q_2=(c/2) \left( q_3-q_1 \right)$ and $q_5=(r/2) \left( q_6-q_4 \right)$.

Already if $A=\beta=0$ but $b$ is not restricted to zero, the pressure becomes non-zero via
\begin{equation}
    p^* = \frac{bc}{2} { \frac{ K_2(r) }{ K_1(c) } } \, \sin (2\theta) \, .
\end{equation}
Unlike the case of translational motion considered previously \cite{daddi2024hydrodynamics}, for which the solution diverges logarithmically as \( c \to 0 \), it remains well-behaved in this limit  for rotational motion.
{
Accordingly, the solution for rotational motion remains well-behaved in the absence of friction.
Moreover, from Eq.~\eqref{eq:lambda_pm}, it follows that for \( A=\beta=b = 0 \) the dimensionless viscosity coefficients become \(\lambda_\pm = 1 \), leading to vanishing pressure. This is also predicted by the Evans--Sackmann theory, for which $v_\rho = 0$ and $v_\theta$ is a function solely of~$\rho$.
}

\begin{table}
    \centering
{\renewcommand{\arraystretch}{2.3}
    \begin{tabular}{|c|c|}
    \hline 
      $g_2$ & $\cfrac{A \left( A-8 \right) w }{64} \left( 1- w^2 \right)^2$ \Tstrut \\
    $g_4$ & $-\cfrac{A^2 w}{128} 
		\left( 1 - w^2 \right)^2 \left( 1-6w^2 + 7w^4  \right)$ \\[3pt]
  \hline
  $h_0$ & $w$  \\
		$h_2$ & $\cfrac{A \left( A-8 \right)w^3 }{64} \left( 1-w^2 \right) $ \\
		$h_4$ & $-\cfrac{A^2 w^3}{128} 
		\left( 1-w^2 \right) \left( 2-8w^2+ 7w^4 \right)$ \\[5pt]
  \hline
  $u_1$ & $\cfrac{b w^2}{128 } 
		\left( A^2-8A+128 \right)$ \Tstrut \\
		$u_2$ & $-\cfrac{A w^2}{4 } \left( 2 -3w^2 \right)$  \\
		$u_3$ & $\cfrac{bA w^2 }{256 } \left( 16-3A \right)\left( 3 -12w^2 + 10w^4 \right) $ \\
		$u_4$ & $-\cfrac{A^2 w^2}{64 }
		\left( 4 - 30w^2 +60w^4 - 35w^6 \right)$  \\
		~~$u_5$~~ & ~~$\cfrac{bA^2 w^2}{256 }
		\left( 5 - 60w^2 + 210w^4 - 280w^6 + 126 w^8 \right)$~~  \\[3pt]
  \hline
    \end{tabular}
    }
    \caption{Expressions for the coefficients defining the Fourier series for the hydrodynamic velocity and pressure fields, see Eqs.~\eqref{eq:v_p_form_Fourier}, up to second order in \( A \) in the case of vanishing friction, that is, for $\alpha_\parallel = \alpha_\perp = 0$. }
    \label{tab:solution-no-firction}
\end{table}

\section{Resistance coefficient}

Thus, in the previous section we derived the leading-order corrections to the hydrodynamic velocity and pressure fields around a thin disk steadily rotating in a homogeneous, weakly uniaxially anisotropic two-dimensional film of fluid under linear friction with its environment. Our next step is to quantify the leading-order corrections in both anisotropy parameters, \(A\) and \(\beta\), to the rotational resistance coefficient of the disk. 

The frictional torque $T$ exerted on the rotating disk can be determined by integrating the moment of the viscous traction over the circumference of the disk~\cite{evans1988translational},
\begin{equation}
    T = a \int_0^{2\pi} -\sigma_{\rho \theta} \bigg|_{\rho = a} \, a \,\mathrm{d} \theta \, .
    \label{eq:TORQUE}
\end{equation}
Defining 
\begin{equation}
    \Pi_{\rho\theta} 
    = v_{\theta, \rho} + \frac{1}{\rho}
    \left( v_{\rho, \theta} - v_\theta \right)  \, , 
\end{equation}
representing the usual component of the viscous stress tensor, it follows that
\begin{equation}
    -\sigma_{\rho\theta} = \nu_3  
   \left( \Pi_{\rho\theta} + \Phi \right) \, ,
\end{equation}
where we have defined
\begin{equation}
    \Phi = \frac{A}{8} \left( 1-\cos(4\theta) \right) \Pi_{\rho\theta}
    + \frac{ v_\rho + v_{\theta, \theta} }{\rho} \, \delta_- - v_{\rho, \rho} \, \delta_+ \, .
\end{equation}
together with the abbreviation
\begin{equation}
    \delta_\pm = \frac{1}{2} \left( \frac{A}{4} \, \sin(4\theta) \pm b \sin (2\theta) \right) \, .
\end{equation}
The quantity \(\Phi\) is related to anisotropy and vanishes for an isotropic fluid of \(A = b = 0\).

We now define the hydrodynamic rotational resistance coefficient $R$, which relates the angular velocity $\Omega$ to the corresponding components of the resistance torque $T$, as $T = -R \Omega$.
Using the  solution in terms of the Fourier series provided in Eq.~\eqref{eq:v_p_form_Fourier}, evaluating it at \(\rho = a\), and considering the facts that \(h_0 = 1\) and \(g_n = h_n = 0\) for \(n \ge 1\), along with the incompressibility equation implying \(\mathrm{d} g_n/\mathrm{d} \rho = 0\) for \(n \ge 1\), all evaluated at \(\rho = a\), it follows that
\begin{equation}
    \frac{R}{\pi a^2 \nu_3} =
    2+\frac{A}{4} -
    \left.
    a \, \frac{\mathrm{d}}{\mathrm{d} \rho} \left( 
    \left( 2+\frac{A}{4} \right) h_0
    -\frac{A}{8} \, h_2 \right)
    \right|_{ \rho=a } .
    \label{eq:drag_final}
\end{equation}

To present the result, we define the abbreviations $\xi_i =  K_i(c)/K_1(c)$. It follows that
\begin{equation}
     \xi_2 = \frac{2}{c} + \xi_0 \,, \qquad
    \xi_3 = 1 + \frac{4}{c} \, \xi_2\, . 
\end{equation}
Using these, we obtain
\begin{align}
	R &= a^2 \, \nu_3  \Big( R^{\mathrm{IES}} + R^{(1,0)} A
	+ R^{(0,1)} \beta \notag \\
	&\quad+ R^{(2,0)} A^2
	+ R^{(1,1)} A\beta
	+ R^{(0,2)} \beta^2 
	\Big) \, .
\end{align}
In this expression, we have defined
\begin{equation}
    R^{\mathrm{IES}} = 2\pi c \, \xi_2 
\end{equation}
to denote the resistance coefficient {for the isotropic Evans--Sackmann model}.
This result is in full agreement with Ref.~\onlinecite{evans1988translational}, which reports the translational and rotational drag coefficients for a disk moving in a fluid membrane supported by a rigid substrate.
Additionally, if \(\alpha_\parallel = \alpha_\perp = 0\), we recover the classical result for the rotational resistance coefficient as given by Eq.~\eqref{eq:first-eqn}.
The leading-order corrections to the resistance coefficient due to the anisotropies in viscosity $A$ and in friction $\beta$ are given by
\begin{subequations}
	\begin{align}
		R^{(1,0)} &= \frac{\pi}{8}
		\Big( 4 + \left( 1-\xi_0^2\right) c^2 \Big) \, , \\[2pt]
		R^{(0,1)} &= \pi c \big( \left( 1-\xi_0^2\right)c-2\xi_0 \big) \, .
	\end{align}
\end{subequations}
In addition, we find the corrections to quadratic order in the parameters of anisotropy as
\begin{subequations}
	\begin{align}
		R^{(2,0)} &= -\frac{\pi }{512}  \,
  \frac{ c^2 }{ \xi_3 } \left( \Pi \, c + \Lambda \right) \, , \\[2pt]
		R^{(1,1)} &= \frac{\pi}{8} \, c^2
		\Big( \xi_0 \left( \xi_0^2-1\right)c+2\xi_0^2-1 \Big) \,  , \\[3pt]
		R^{(0,2)} &= \frac{\pi}{4} \, c \,
		\Big( 3\xi_0 \left( \xi_0^2-1\right)c^2+ \left( 8\xi_0^2-5\right)c + 4\xi_0 \Big) \, , 
	\end{align}
\end{subequations}
where we have introduced the abbreviations
\begin{subequations}
    \begin{align}
    \Pi &= 14\xi_2^3+4\xi_0^2\xi_2-24\xi_0\xi_2^2-\xi_0+7\xi_2 \,, \\[.1cm]
    \Lambda &= 8\xi_2^2 \left( \xi_0^2-4\xi_0\xi_2+\xi_2^2 \right) .
\end{align}
\end{subequations}

The hydrodynamic rotational mobility, that is, the inverse of the resistance coefficient, is similarly expressed as a power series in $A$ and $\beta$. It can be written in the form
\begin{align}
    \mu &= \frac{ \mu^{\mathrm{IES}} }{ \nu_3 }
    \Big( 1 + \mu^{(1,0)} A + \mu^{(0,1)} \beta \notag \\
    &\quad+ \mu^{(2,0)} A^2 + \mu^{(1,1)} A\beta + \mu^{(0,2)} \beta^2
    \Big) \, , 
\end{align}
where $\mu^{\mathrm{IES}} = 1/R^{\mathrm{IES}}$ represents the rotational mobility {for the isotropic Evans--Sackmann model}.
The correction terms arising from the anisotropies in viscosity and friction are derived to leading order in $A$ and $\beta$ as
\begin{subequations}
    \begin{align}
    \mu^{(1,0)} &= -\frac{ R^{(1,0)} }{ R^{\mathrm{IES}} } \, , \\[2pt]
    \mu^{(0,1)} &= -\frac{ R^{(0,1)} }{ R^{\mathrm{IES}} } \, .
\end{align}
\end{subequations}
Moreover, the quadratic corrections can then be expressed as 
\begin{subequations}
    \begin{align}
    \mu^{(2,0)} &= \left( \mu^{(1,0)} \right)^2 
    - \frac{ R^{(2,0)} }{ R^{\mathrm{IES}} } \, , \\[2pt]
    \mu^{(1,1)} &= 2 \mu^{(1,0)} \mu^{(0,1)}
    - \frac{ R^{(1,1)} }{ R^{\mathrm{IES}} } \, , \\[2pt]
    \mu^{(0,2)} &= \left( \mu^{(0,1)} \right)^2 
    - \frac{ R^{(0,2)} }{ R^{\mathrm{IES}} } \, .
\end{align}
\end{subequations}
Besides, we remark that, in the limit of vanishing $c=\alpha_{\parallel}a$, the rotational mobility up to quadratic order in the parameters of anisotropy is found as
\begin{equation}
    \lim_{c\to 0} \mu = \frac{1}{4\pi\nu_3}
    \left( 1 - \frac{A}{8} + \frac{3 A^2}{128} \right) \, .
    \label{eq:mobi_lim_c0}
\end{equation}

\begin{figure}
    \centering
    \includegraphics[scale=0.85]{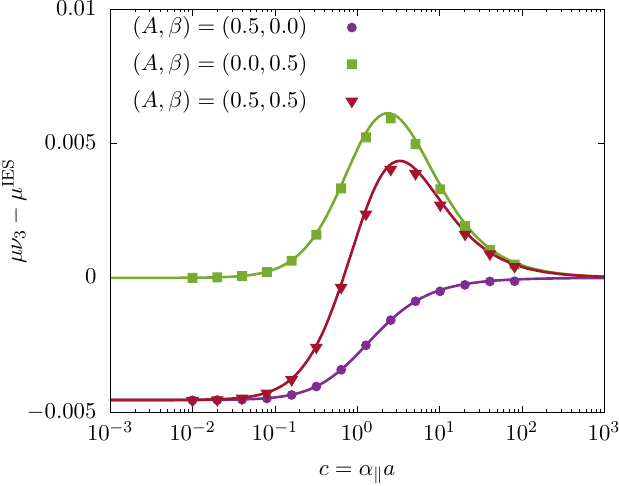}
    \caption{Comparison of the correction to the scaled rotational mobility function against the scaled friction coefficient for various combinations of viscosity and friction anisotropy, as obtained from fully resolved finite-element simulations (symbols) and perturbative analytical theory (lines).}
    \label{fig:mobirot}
\end{figure}

In Fig.~\ref{fig:mobirot} we present a comparison of the correction to the scaled rotational mobility function for different combinations of viscosity and friction anisotropy, derived from  our perturbative analytical theory (lines) and fully resolved finite-element simulations (symbols), see Sec.~\ref{sec:fem}. 
Results are plotted against the scaled friction coefficient {\(c = \alpha_\parallel a\)}. Generally, the corrections are very small compared to the bulk isotropic mobility \(\mu^{\mathrm{IES}}\), making them challenging to determine in real situations.
For \((A,\beta)=(0.5, 0)\), representing the case of viscosity anisotropy only, the correction reaches its maximum value as \(c \to 0\), exhibiting a logistic-like evolution, and vanishes as \(c \to \infty\), corresponding to the limit of infinite friction.
For \((A,\beta)=(0, 0.5)\), representing the case of friction anisotropy only, the curve exhibits a bell-shaped form, peaking around \(c \sim 1\). In this case, the correction is positive across the entire range of \(c\), indicating that friction anisotropy enhances the rotational motion of the disk.
By combining both modes of anisotropy, for \((A,\beta)=(0.5, 0.5)\), an interesting evolution is observed, showcasing the complex interplay between viscosity anisotropy and friction anisotropy.
Overall, there is excellent agreement between the theoretical predictions and the simulation results.

{
We remark that in an infinitely extended, isotropic two-dimensional fluid medium, there is no coupling between translational and rotational components in the resistance matrix of a moving disk. Similarly, we do not find such coupling for a disk translating in a uniaxially anisotropic fluid. That is, the translation of the disk does not induce rotational motion and vice versa due to this lack of coupling.
}

\section{Exact solution in the absence of friction}
\label{sec:exact}

As mentioned previously, in the specific case of vanishing friction with the environment, that is, for $\alpha_\parallel = \alpha_\perp = 0$, the solution for rotational motion of the disk remains convergent in two dimensions. Both the rotational mobility of the disk as well as the hydrodynamic velocity and pressure fields remain well-behaved. 
This is at odds with the corresponding two-dimensional situation of translational motion \cite{daddi2024hydrodynamics}. 
We thus analyze further the situation of rotational motion under vanishing linear friction. 

In this scenario, the integrals $\Gamma_{mn}$, defined by Eq.~\eqref{eq:GAMMA}, simplify significantly and can be expressed as
\begin{equation}
    \Gamma_{mn} = \frac{(-1)^{m+n}}{2B}
    \frac{ \Gamma(m+n+1) }{ \Gamma(m-n+1) \Gamma(2n+2) } \, \varphi_{mn} \, , 
    \label{eq:GAMMA-no-friction}
\end{equation}
wherein $\Gamma$ denotes the Euler Gamma function and $_2F_1$ the Gaussian
hypergeometric function.
Moreover,
\begin{equation}
    \varphi_{mn} = w^{2n+1} 
    {}_2F_1 \left( n-m,n+m+1; 2n+2; w^2 \right) , 
\end{equation}
which equals zero when \( n > m \).
Still, $\Gamma_{mn}$ depends on~$\phi$ because $B$ is a function of~$\phi$.

The solution for the force density, described by Eq.~\eqref{eq:Kraft-distr-simplified}, simplifies to only include the term for \( n = 0 \). 
We find that along the circumference of the disk, where $w=1$,
\begin{equation}
    \Gamma_{m,n} (w=1) = \frac{ \delta_{mn} }{2 B \left( 2m+1\right)} \, .
\end{equation}
To satisfy the boundary conditions at the circumference of the disk, it is necessary that \( d_1 = -c_1 \).
Accordingly, Eq.~\eqref{eq:GAMMA-no-friction} for \( n=0 \) evaluates to
\begin{equation}
    \Gamma_{m\, 0} = \frac{ (-1)^{m} }{B} \, W_m \, ,
\end{equation}
where
\begin{equation}
    W_m = w \, {}_2 F_1 \left( -m, m+1; 2; w^2 \right) \, .
\end{equation}
We find
\begin{equation}
    c_1 = \left( 1+\frac{A}{4} \right)^\frac{1}{2}  \, .
\end{equation}

The radial and azimuthal components of the fluid velocity, see Eqs.~\eqref{eq:v_p_form_Fourier},  can be written in the form
\begin{subequations}
    \begin{align}
    v_\rho &= a\Omega \,
    \sum_{m=0}^\infty \, W_m
    \int_0^{2\pi}
    Q_m( \theta, \phi) 
    \sin ( \theta-\phi ) \,
    \mathrm{d}\phi \, , \\[3pt]
    v_\theta &= a\Omega \,
    \sum_{m=0}^\infty \, W_m
    \int_0^{2\pi}
    Q_m( \theta, \phi) 
    \cos ( \theta-\phi ) \,
    \mathrm{d}\phi \, , 
\end{align}
\end{subequations}
where we have defined
\begin{equation}
    Q_m = \frac{(-1)^{m}}{\pi B}\, \left( A+4 \right)^{\frac{1}{2}} \cos \left( (2m+1) (\theta-\phi) \right) ,
\end{equation}
bearing in mind that $B$ is itself a function of~$\phi$.
Defining 
\begin{equation}
    \chi_n = \left( \frac{ A^\frac{1}{2} }{2+ \left(A+4\right)^\frac{1}{2} } \right)^n \, , 
\end{equation}
we find after some algebra that both \( g_n \) and \( h_n \) vanish if \( n \) is odd. For \( n \) even, we obtain for \( n \geq 2 \)
\begin{subequations}
    \begin{align}
    g_n &= -2 \chi_n
    w \, {}_2 F_1 \left( -n,n; 2; w^2 \right), \label{eq:g_n} \\
    h_n &= -n\chi_n
    w^3 {}_2F_1 \left( n+1,-n+1; 3, w^2 \right) . \label{eq:h_n}
\end{align}
\end{subequations}
In addition, $g_0 = 0$ and $h_0 = w$.

The expression for the pressure, see Eqs.~\eqref{eq:v_p_form_Fourier}, can similarly be derived and expressed as an infinite series, resulting in the series coefficients \( u_n \). 
We obtain, after some algebra,
\begin{equation}
    u_n = 2n \chi_n E_n w^2
    {}_2F_1 \left( n+1, -n+1; 2; w^2 \right) \, , 
    \label{eq:u_n}
\end{equation}
wherein
\begin{equation}
    E_n = 
    \begin{cases}
    2b A^{-\frac{1}{2}} & \text{if} \quad n = 2k+1 \,, \\
    -\left( A+4\right)^\frac{1}{2} & \text{if} \quad n = 2k\,.
\end{cases} \label{eq:E_n}
\end{equation}
Expanding the final results in a power series of \( A \), we confirm that the expressions given in Table~\ref{tab:solution-no-firction} are exactly recovered.

The hydrodynamic rotational resistance coefficient can be calculated by substituting the expressions for \(h_0\) and \(h_2\) from Eq.~\eqref{eq:h_n} into Eq.~\eqref{eq:drag_final}.
It can be precisely calculated as
\begin{equation}
    R = 2\pi \nu_3 \left( A+4\right)^\frac{1}{2} \, . \label{eq:R}
\end{equation}
When we expand the inverse of this result in a power series of~$A$, we successfully recover the expression for the mobility function given by Eq.~\eqref{eq:mobi_lim_c0}.

{Notably, Eq.~\eqref{eq:R} shows that \(R = 0\) when \(A = -4\), which is the smallest value \(A\) can take when requiring both physically and mathematically that \(B \ge 0\). While providing a physical interpretation might be challenging, it is not surprising that the case \(A = -4\) is special, given that it represents the lower bound of this interval of~$A$.}

Next, we evaluate the pressure field at the circumference of the disk at \(\rho = a\). Then, the infinite sum can be computed analytically. To this end, we utilize
\begin{subequations}
    \begin{align}
    {}_2F_1 \left( -2k,2k+2; 2; 1 \right) = \frac{1}{2k+1} \, , \\[3pt]
    {}_2F_1 \left( 2k+1,-2k+1; 2; 1 \right) = -\frac{1}{2k} \, .
\end{align} \label{eq:needed-for-pressure}
\end{subequations}

By utilizing the Fourier series representation provided in Eq.~\eqref{eq:p_form_Fourier} for the pressure field, along with Eqs.~\eqref{eq:u_n} and \eqref{eq:needed-for-pressure}, the resulting infinite series expansion can be precisely evaluated.
It is noteworthy that the expression for the general term in the series takes a specific form depending on the parity of \( n \); refer to Eq.~\eqref{eq:E_n}.
After some calculation, we ultimately find 
\begin{equation}
    p(\rho=a,\theta) = \frac{ \Omega \nu_3 }{a}
    \left( A+4\right)^\frac{1}{2}
    \frac{ 4b \sin(2\theta) + A\sin(4\theta) }{ 8 + A \left( 1-\cos(4\theta)\right) } \, .
\end{equation}

Since the computation of the hypergeometric function in some software requires additional toolboxes, we provide the polynomial representation in the form of a terminating series expansion. This termination occurs because the first or second argument is a non-positive integer.
We obtain 
\begin{align}
    {}_2F_1 ( -n,n &; 2; w^2 ) 
     \notag \\
    &=\sum_{k=0}^n (-1)^k
    \binom{n}{k} \frac{ (n)_k }{(2)_k} \, w^{2k} \notag \\
    &= \sum_{k=0}^n
    (-1)^k \frac{n(n+k-1)!}{k!^2 (k+1)(n-k)!} \, w^{2k} \, ,
\end{align}
which is involved in the calculation of \(g_n\), see Eq.~\eqref{eq:g_n}.
Here $(q)_n$ is the (rising) Pochhammer symbol, defined by
\begin{equation}
    (q)_n = \begin{cases}  1  & n = 0\,,\\
  q(q+1) \cdots (q+n-1) & n > 0\,.
 \end{cases}
\end{equation}
In addition,
\begin{align}
    {}_2F_1 ( n+1 & ,-n+1 ; 3; w^2 ) \notag \\
    &=\sum_{k=0}^{n-1} (-1)^k
    \binom{n-1}{k} \frac{ (n+1)_k }{(3)_k} \, w^{2k} \notag \\
    &= \sum_{k=0}^{n-1}
    (-1)^k \frac{2(n+k)!}{n k! (k+2)! (n-k-1)!} \, w^{2k} \, ,
\end{align}
which enters the calculation of \(h_n\), see Eq.~\eqref{eq:h_n}.
The series for \( {}_2F_1(n+1, -n+1; 2; w^2) \), which is relevant for calculating the pressure field; see Eq.~\eqref{eq:u_n}, can similarly be obtained by noting that
\begin{equation}
    (3)_k = \left( 1+\frac{k}{2} \right) (2)_k \, .
\end{equation}

Since the numerical computation of the series describing the hydrodynamic fields requires truncation once a certain precision is reached, it is essential to study the convergence of the series and estimate the number of terms needed to achieve the desired precision.
This convergence study is detailed in the Appendix.

\begin{figure*}
    \centering
    \includegraphics[scale=0.5]{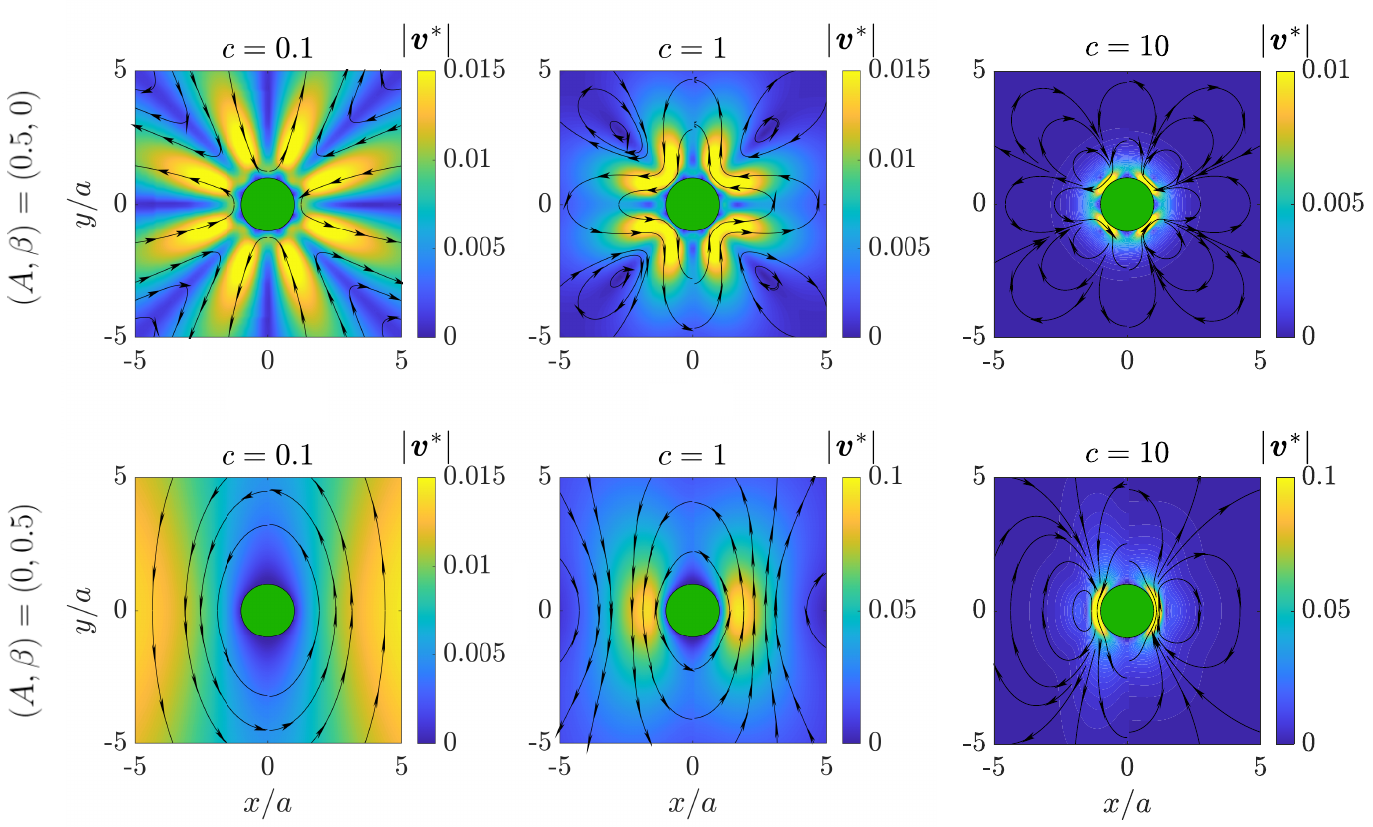}
    \caption{
    Streamlines and quiver plots of the {leading-order correction to the flow field due to uniaxial anisotropy in  viscosity and friction} \(\bm{v}^*\) for a disk rotating steadily in a two-dimensional, uniaxially anisotropic fluid. The results are shown for three values of the parameter \(c = \alpha_\parallel a\), namely \(c = 0.1\) (left column), \(c = 1\) (middle column), and \(c = 10\) (right column). 
    Additionally, two combinations of the parameters \(A\) and \(\beta\) quantifying the anisotropy of viscosity and friction  are examined, namely \((A, \beta) = (0.5, 0)\) (top row) and \((A, \beta) = (0, 0.5)\) (bottom row). 
    {Perturbative solutions} are depicted on the right-hand sides (\(x > 0\)), while results from finite-element simulations are shown on the left-hand sides (\(x < 0\)).
    We recall that \(\bm{v}^*\) is rescaled by \(a\Omega\) and that \(c = \alpha_\parallel a\).
    }
    \label{fig:flow}
\end{figure*}

\begin{figure}
    \centering
    \includegraphics[scale=0.6]{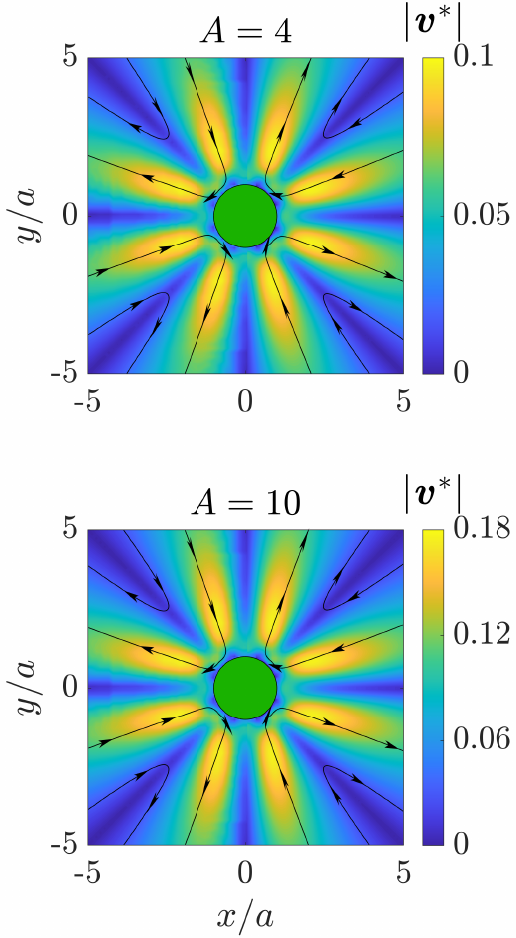}
    \caption{Streamlines and quiver plots of the {leading-order correction to the flow field due to uniaxial viscosity anisotropy} \(\bm{v}^*\) for a disk rotating steadily in a two-dimensional, uniaxially anisotropic fluid in the absence of friction, that is, for $\alpha_\parallel = \alpha_\perp = 0$.
    Exact analytical solutions are shown on the right-hand sides (\(x > 0\)), while results from finite-element simulations are shown on the left-hand sides (\(x < 0\)).
    Again, \(\bm{v}^*\) is rescaled by \(a\Omega\).
    }
    \label{fig:flowexact}
\end{figure}

\section{Comparison with results from finite-element simulations}
\label{sec:fem}

To verify the analytical findings, we conduct numerical simulations employing the Gascoigne 3D finite element library~\cite{Braack2021}. 
Following a similar methodology as described before~\cite{daddi2024hydrodynamics}, we formulate the governing equations of fluid motion in a variational framework. 
Both velocity and pressure fields are discretized using quadratic finite elements.

To address the open boundary conditions, we define a sufficiently large numerical domain within which the solution decays adequately towards the boundaries. 
We optimize computational efficiency by implementing local mesh refinement around the disk. 
The resulting linear systems are solved using the well-established GMRES solver~\cite{saad86}, preconditioned with a parallel multigrid approach~\cite{FailerRichter2019}. 
Typical computation times are just a few seconds per configuration.

In Fig.~\ref{fig:flow} we present streamlines and quiver plots of the {leading-order correction to the flow field} for a disk undergoing steady rotational motion in a two-dimensional, uniaxially anisotropic fluid.
The results are depicted for three values of the parameter \(c = \alpha_\parallel a\), namely \(c = 0.1\) (left column), \(c = 1\) (middle column), and \(c = 10\) (right column).
We examine two combinations of the parameters quantifying the anisotropy of viscosity and friction, \(A\) and \(\beta\), namely \((A, \beta) = (0.5, 0)\) (top row) and \((A, \beta) = (0, 0.5)\) (bottom row). 
{Perturbative  results} are displayed in the range $x > 0$, while results from finite-element simulations are shown in the range $x < 0$.

{
In the case of vanishing anisotropy of friction, we have demonstrated that \( g_n \) and \( h_n \), describing the radial and azimuthal velocities, respectively, vanish for odd \( n \). Consequently, \( v_\rho(\rho, \theta + \pi/2) = v_\rho(\rho, \theta) \) and \( v_\theta(\rho, \theta + \pi/2) = v_\theta(\rho, \theta) \), indicating that the velocity remains unchanged upon rotations of \(\pi/2\).
Anisotropy of viscosity does not notably distort the flow profile with respect to the axis of anisotropy \( \bm{\hat{n}}\|\bm{\hat{x}} \), see the top row of Fig.~\ref{fig:flow}.
We further observe non-monotonic behavior in the magnitude of velocity, which seems to occur in all situations.
This is because the correction \(\bm{v}^*\) to the flow field vanishes at the circumference of the disk and at infinity. If \(\bm{v}^*\) does not vanish overall, its magnitude must thus reach a maximum value somewhere in between.
}

In Fig.~\ref{fig:flowexact}, we present the {leading-order correction to the flow field} for rotational motion in a nematic liquid crystal without friction, a scenario for which exact analytical expressions can be derived, see Sec.~\ref{sec:exact}. Results are illustrated for the two values of \(A=4\) and \(10\). They are sufficiently large for the perturbative solution not to remain quantitatively correct any more in general. The analytical solution, displayed on the right-hand sides, aligns perfectly with the finite-element simulations shown on the left-hand sides. Notably, the flow field features regions of large magnitudes of velocity. These regions are arranged along the edge of an octagon near the surface of the disk.

In the listed contexts, we concentrate on presenting the leading-order correction to the flow field due to uniaxial viscosity and friction anisotropy rather than the entire flow field. In this way, we enhance the visualization of the impact of the anisotropies. 
Our results reveal that the flow field is significantly influenced by changes in the parameters of anisotropy. In some cases, they exhibit vortical structures with closed streamlines and regions of high magnitude of velocity near the disk.
Overall, our theoretical predictions based on a perturbative approach align very well with results from finite-element simulations that solve the complete set of hydrodynamic equations.

\section{Conclusions}
\label{sec:conclusions}

In summary, we consider the situation of a rotating disk in a two-dimensional, uniaxially anisotropic, incompressible fluid film or membrane under low-Reynolds-number conditions. The anisotropy direction is globally aligned. Additional linear friction may occur, for example, with a supporting substrate or other top and bottom plates confining the thin fluid medium. We consider a rotating disk laterally embedded in this fluid film under no-slip surface conditions along its circumference. 

Following an expansion into two parameters of anisotropy, one related to viscosity and the other to the coefficients of linear friction, we derive analytical expressions up to quadratic order in the anisotropies for the hydrodynamic flow and pressure fields as well as for the resistance and mobility coefficients of the rotating disk. Our results compare well with those obtained from finite-element simulations solving the original equations of low-Reynolds-number hydrodynamics in the presence of the rotating disk. 

From a conceptual point of view, it is interesting to note that the solution remains well-behaved even in the limit of vanishing additional linear friction. This is in contrast to the case of translational motion of the disk~\cite{daddi2024hydrodynamics}. Yet, it is in line with the situation for isotropic two-dimensional media, where likewise the solution diverges for translational motion but remains well-defined under rotations of the disk \cite{richter22}. Formally, this feature is related to a logarithmic divergence emerging in the two-dimensional Green's function that relates the stimulus by a net force to the induced flow field. 

Various scenarios of practical application of our results are conceivable. For example, at some point stirring in flat cells of nematic liquid crystals \cite{degennes95} could be desired. Purposes may include mixing in such cells or flow-induced restructuring upon significant misalignment. To this end, flat disk-shaped magnetic nano- or microplatelets could be inserted. They can be rotated from outside by external magnetic fields, if their magnetic easy axis is oriented in-plane \cite{li2004microstructure, beek2006magnetic, beek2008influence, reich2010sedimentation, daddi2020dynamics} and is sufficiently anchored along a specific in-plane direction so that induced rotation can be achieved. 

During the further course of the topic and within a wider context, expansions of the theory to thin elastic membranes functionalized by rigid, plate-like inclusions represent a next step~\cite{puljiz2017forces, lutz2024internal}. 
We plan to extend our considerations to elastic composite systems in the future, particularly aiming at a theoretical consideration of membranes of globally aligned liquid crystal elastomers \cite{kupfer1991nematic, urayama2007selected, menzel2007nonlinear, menzel2009response, fleischmann2012one, lagerwall2023liquid}. 

{
Finally, in a two-dimensional uniaxially anisotropic fluid film, a disk undergoing rigid rotation would potentially affect the alignment of the axis of anisotropy.
We here have considered a simplified and idealized situation, in which the nematic director is assumed to be permanently aligned along a specific direction. This may be due to, for instance, a remotely exerted electric or magnetic field. Our assumption allows the hydrodynamic equations of fluid motion to be considered in a well-addressable form, enabling analytical progress and occasionally exact solutions.
Relaxing this assumption to consider a more general scenario, in which the director field is coupled to the velocity field, is a challenging endeavor for the future. As a first step, the situation of a slowly varying director field is particularly interesting, as it may allow for a solution based on perturbative approaches.} %These calculations are indeed of great interest and could be explored in detail in a future publication.

\section*{Data availability}

The Maple~24 scripts containing the final expressions for the velocity field are accessible through the Zenodo repository using the DOI \href{https://zenodo.org/records/12721219}{10.5281/zenodo.12721219}.
The data and {numerical codes} that support the findings of this study are available from the corresponding author upon reasonable request.

\section*{Author contribution}

A.D.M.I. and A.M.M. designed the research and wrote the manuscript. A.D.M.I. performed the analytical calculations. T.R. conducted the finite-element simulations. A.D.M.I., E.T., M.P., T.R., and A.M.M. reviewed, edited, and provided feedback on the manuscript.

\begin{acknowledgments}
E.T.\ acknowledges the support received from the EPSRC under grant no.\ EP/W027194/1. A.M.M.\ acknowledges support provided by the Deutsche Forschungsgemeinschaft (DFG, German Research Foundation) through research grant no.\ ME 3571/5-1. Furthermore, A.M.M.\ thanks the DFG for support via the Heisenberg grant no.\ ME 3571/4-1.
\end{acknowledgments}

\appendix

\section{Convergence study of the series functions}

In this Appendix, we provide a comprehensive convergence study of the series function defined by Eqs.~\eqref{eq:v_p_form_Fourier}, which characterizes the hydrodynamic fields as derived in Sec.~\ref{sec:exact}, specifically in the scenario without friction.

For a given series function of the form
\begin{equation}
    \sum_{n=0}^\infty f_n \, \Theta (2n\theta) \, , 
\end{equation}
where \(\Theta\) is either a sine or a cosine function, we define the truncation error for a sufficiently large integer~\( N \) as
\begin{equation}
    \mathcal{E}_M = \left| \sum_{n=N}^\infty f_n \, \Theta (2n\theta) \right|
    \le \sum_{n=N}^\infty |f_n|  \, .
\end{equation}
To achieve a specified absolute error \( \epsilon \), we must ensure that \( \mathcal{E}_M \) remains below this \( \epsilon \) threshold.
Without loss of generality, we assume that \(N=2M\) is an even number.

The hypergeometric functions that appear in the definitions of \(g_n\), \(h_n\), and \(u_n\) reach their maximum value of 1 at \(w=0\) within the interval \([0,1]\).
Defining 
\begin{equation}
    r = \frac{|A|}{ \left( 2+(A+4)^\frac{1}{2} \right)^2 } \,\, \in [0,1] \, , 
    \label{eq:def_r}
\end{equation}
it follows that 
\begin{equation}
    \sum_{n=N}^\infty |g_n| 
    \le \sum_{k = M }^\infty 2 r^k  
    = \frac{2 r^M}{1 - r} \, , \label{eq:g_n_sum}
\end{equation}
where we have used the change of index $n=2k$.
Solving the inequality 
\begin{equation}
    \frac{2 r^M}{1-r} < \epsilon 
\end{equation}
for $M$ we obtain
\begin{equation}
    M \ge \left\lceil \frac{ \ln \left( \frac{\epsilon}{2} (1-r) \right) }{\ln r} \right\rceil \, ,
    \label{eq:Nbr_terms_gn}
\end{equation}
with $\lceil x \rceil$ representing the ceiling of a real number $x$, defined as the smallest integer greater than or equal to $x$.

In addition, we have
\begin{equation}
    \sum_{n=N}^\infty |h_n| 
    \le \sum_{k = M }^\infty 2 k r^k  
    = \frac{2 r^M}{1-r} \left( M + \frac{r}{1-r} \right) \, , 
    \label{eq:h_n_sum}
\end{equation}
where again we have set $n=2k$.
Neglecting, for simplicity, the ratio \( r/(1-r) \) in the brackets against \( M \) in Eq.~\eqref{eq:h_n_sum}, we then need to solve for $M$ the inequality
\begin{equation}
    \frac{2M r^M}{1-r} < \epsilon 
\end{equation}
to get
\begin{equation}
    M \ge \left\lceil \frac{ W_{-1} \left( \frac{\epsilon}{2}(1-r)\ln r \right) }{\ln r} \right\rceil \, , 
    \label{eq:Nbr_terms_hn}
\end{equation}
with $W_{-1}$ representing a branch of the Lambert $W$ function.

\begin{figure}
    \centering
    \includegraphics[scale=0.8]{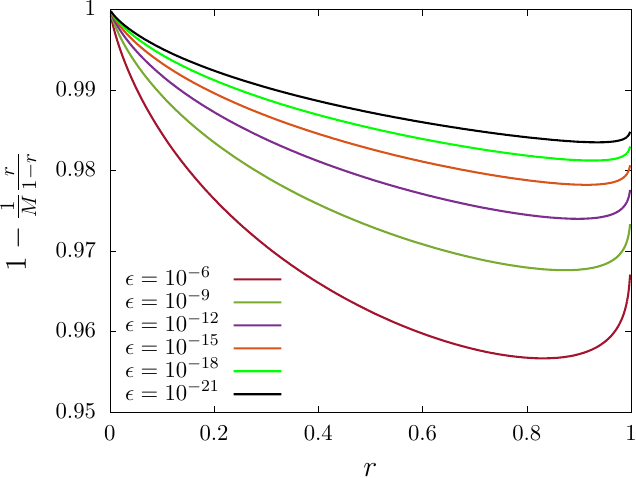}
    \caption{Illustration of the validity of the approximation in Eq.~\eqref{eq:h_n_sum} of neglecting the ratio $r/(1-r)$ when compared to $M$, as described by Eq.~\eqref{eq:Nbr_terms_hn}, for various values of the absolute error $\epsilon$.
    Here, we used the value of \( M \) without applying the ceiling function to prevent a zigzag pattern.}
    \label{fig:error}
\end{figure}

In Fig.~\ref{fig:error} we show an illustration demonstrating the validity of the approximation in Eq.~\eqref{eq:h_n_sum} by neglecting the ratio \( r/(1-r) \) relative to \( M \), as described in Eq.~\eqref{eq:Nbr_terms_hn}, for various values of the absolute error \( \epsilon \).
The plotted function approaches one as \( r \) tends toward 0 or 1, and reaches its minimum value at an intermediate point between these two extremes.
It is evident that this approximation is well justified, as the ratio remains less than 5~\% 
across the entire range of \( r \) for \( \epsilon = 10^{-6} \) and decreases to less than 2~\% for \( \epsilon = 10^{-15} \).

\bgroup
\def\arraystretch{1.75} % 2.5
    \begin{table}
    \centering
\begin{tabular}{|c|ccc|ccc|ccc|}
    \hline
     \multirow{2}{*}{$\epsilon$} & \multicolumn{3}{c|}{$g_n$} & \multicolumn{3}{c|}{$h_n$} & \multicolumn{3}{c|}{$u_n$}  \\
    \cline{2-10}
     & ~~A~~ & ~~B~~ & ~~C~~ & ~~A~~ & ~~B~~ & ~~C~~ & ~~A~~ & ~~B~~ & ~~C~~  \\
    \hline
    \hline
    $10^{-6}$ & 7 & 22 & 160  & 8 & 27 & 211  & 9 & 31  & 254  \\
    \hline
    $10^{-9}$ & 10 & 32 & 226  &  11 & 38 & 279  &  12 & 41  & 322 \\
    \hline
    $10^{-12}$ & 13 & 42 & 291  &  14 & 48 & 347  &  15 & 52  & 389  \\
    \hline
\end{tabular}
\caption{The approximate number of terms needed to evaluate the infinite series representing the hydrodynamic velocity and pressure fields, see Eqs.~\eqref{eq:v_p_form_Fourier}. We consider three truncation errors $\epsilon$ and three parameters \( r \), see Eq.~\eqref{eq:def_r}, according to the scheme A, B, and C, corresponding to \( r = 0.1 \), \( r = 0.5 \), and \( r = 0.9 \), respectively. 
}
        \label{tab:error}
    \end{table}
\egroup

For the coefficients \( u_n \) setting the series function of the pressure field, we need to consider terms with both odd and even indices.
We have
\begin{align}
    \sum_{n=N}^\infty |u_n|
    &\le 
    4 (A+4)^\frac{1}{2} \sum_{k=M} k r^k \notag \\
    &\quad+ 
    4b \left| \frac{r}{A} \right|^\frac{1}{2}
    \sum_{k=M} (2k+1) r^k < \epsilon \, .
\end{align}
We obtain
\begin{equation}
    M \ge \left\lceil \frac{ W_{-1} \left( \frac{\epsilon (1-r)}{4 (2+\xi)}\ln r \right) }{\ln r} \right\rceil \, , 
    \label{eq:Nbr_terms_un}
\end{equation}
where
\begin{equation}
    \xi = \left( A+2b \right) \left| \frac{r}{A} \right|^\frac{1}{2} \, .
\end{equation}

In Table~\ref{tab:error}, we present the number of terms required for computing the series defined by \( g_n \), \( h_n \), and \( u_n \), as predicted by Eqs.~\eqref{eq:Nbr_terms_gn}, \eqref{eq:Nbr_terms_hn}, and \eqref{eq:Nbr_terms_un}, respectively. We provide results for three sets of parameters labeled as A, B, and C, corresponding to \( r = 0.1 \), \( r = 0.5 \), and \( r = 0.9 \), respectively. For the computation of \( u_n \), we set \( b = 1 \).
Clearly, a smaller error requires a larger number of terms to achieve the desired precision. Additionally, the number of terms required increases rapidly as \( r \) approaches one. 
A general trend observed is that computing the series defined by \( g_n \) requires fewer terms than that defined by \( h_n \), and \( h_n \) itself requires fewer terms than the series defined by \( u_n \).

%\bibliography{biblio}

\end{document}